\documentclass[twocolumn]{article}
\usepackage{amssymb}
\usepackage{float}
\usepackage{graphics}
\usepackage{abstract}
\usepackage{cite}

\begin{document}

\title{Conditions for plasma evolution to the\ strong general Woltjer-Taylor
state \thanks{{\small Work supported by the National Natural Science
Foundation of China under grant 11105133 and 11175113}}}
\author{Jun-hua Chen$^{1}$ \thanks{
To whom correspondence should be addressed. Email: cjh@ustc.edu.cn} and
Hong-yi Fan$^{1}$ \\
$^{1}${\small Department of Material Science and Engineering, }\\
{\small University of Science and Technology of China,}\\
{\small Hefei, Anhui, 230026, China}}

\twocolumn[

\maketitle

\begin{onecolabstract}

We find that the proof in the recent paper$\textsuperscript{\cite{14}}$ can not justify the authors' conclusion. We provide a real proof that any state will eventually evolves to the Woltjer-Taylor state exponentially. However, this kind of evolution is is mainly due to Joule heat, which also makes the magnetic field vanishes exponentially. Zero Woltjer-Taylor states are not physically attractive. Instead of examine $\Delta$, we introduce the quantity $\theta _{\nabla \times \vec{B},\vec{B}}$ and $R$ to examine if the plasma reaches to the strong (general) Woltjer-Taylor state,
and then derive the condition for the evolution to the strong/general
Woltjer-Taylor state.

$\bigskip $$\bigskip $$\bigskip $$\bigskip $$\bigskip $PACS: 52.30.Cv, 52.55.Lf, 52.55.Tn

\end{onecolabstract}
] \saythanks

\section{Introduction}

Based on the strong astrophysical and laboratory evidence that plasmas tend
to evolve towards the Woltjer-Taylor state (WTS) $\textsuperscript{%
\cite{1,2,3,4,5,6,7,8,9,10}}$, characterized by $\nabla \times \vec{B}%
=\alpha \vec{B}$, and being unsatisfied with Taylor's theory (conjecture) $%
\textsuperscript{\cite{11,12,13}}$, Qin et al $\textsuperscript{\cite{14}}$
developed a theory which they claimed in favor of the evolution towards to
the WTS. They introduced $\Delta \equiv QW-H^{2}\geqslant 0$ for measuring
the deviation of the plasma to the WTS, where $W=$ $\int_{V}\vec{B}%
^{2}d^{3}x $, $H=\int_{V}\vec{A}\cdot \vec{B}Bd^{3}x$, $Q$\ $=\int_{V}\vec{A}%
^{2}d^{3}x$, and $\vec{B}$ is the magnetic field, $\vec{A}$ is the vector
potential. They pointed out that equality $\Delta =0$ holds if and only if $%
\vec{B}=\alpha \vec{A}$ for some constant $\alpha $ and therefore $\nabla
\times \vec{B}=\alpha \vec{B}$. Qin et al\ then proved that $\frac{d}{dt}%
\Delta \leqslant 0$ and $\frac{d}{dt}\Delta =0$ if and only if $\Delta =0$.
Based on the above inferences they drew their conclusion that $%
\lim\limits_{t\rightarrow +\infty }\Delta \left( t\right) =0 $ and the WTS
is reached.

However, the following counterexample shows that Qin's conclusion is not
justified by their argument if there is no further improvement on the
estimation of $\frac{d}{dt}\Delta $. Let $x\left( t\right) \geqslant 0$ be a
function satisfying $\frac{dx}{dt}=-\frac{x}{1+t^{2}}$. We see that $\frac{dx%
}{dt}\leqslant 0$, and $\frac{dx}{dt}=0$ if and only if $x=0$. The solution
to the equation $\frac{dx}{dt}=-\frac{x}{1+t^{2}}$ is $x\left( t\right)
=x\left( 0\right) \exp \left( -\arctan t\right) $. Though $\frac{dx}{dt}<0$
for all $t$, we still have $x\left( +\infty \right) =x\left( 0\right) \exp
\left( -\frac{\pi }{2}\right) >0$.

Further, even if one do manage to prove that $\lim\limits_{t\rightarrow
+\infty }\Delta \left( t\right) =0$, since $\Delta $ is quartic in $\vec{B}$%
, $\Delta ^{\prime }$s approaching zero could possibly be the simple
result of the vanishing of $\vec{B}$, as we will see later in this paper.
Instead of examining $\Delta $ as Qin et al did, we introduce a
dimensionless quantity, i.e., the angle $\theta _{\nabla \times \vec{B},\vec{%
B}}$ between two fields $\vec{B}$ and $\nabla \times \vec{B}$, which is
defined by%
\begin{equation}
\cos \theta _{\nabla \times \vec{B},\vec{B}}=\frac{\left\langle \nabla
\times \vec{B},\vec{B}\right\rangle }{\sqrt{\left\langle \nabla \times \vec{B%
},\nabla \times \vec{B}\right\rangle \left\langle \vec{B},\vec{B}%
\right\rangle }}.  \label{1}
\end{equation}%
As one can see, $\theta _{\nabla \times \vec{B},\vec{B}}$ measures the
deviation of plasma to the strong WTS ( by "strong" we mean the case of $%
\nabla \times \vec{B}=\alpha \vec{B}$ and $\vec{B}\neq 0$, $\alpha $ is a
constant ). Further, if $\alpha \left( \vec{x}\right) $ in equation $\nabla
\times \vec{B}=\alpha \vec{B}$\ varies in space, we say the plasma is in
general WTS. In this case we can introduce another dimensionless quantity,
the relative residual error $R$ (see Eq. (\ref{21})). As we shall explain
below Eq. (\ref{21}), $R$ measures the deviation of plasma to the strong
general WTS. By calculating the evolution of $\theta _{\nabla \times \vec{B},%
\vec{B}}$ and $R$, we can check whether plasma evolves to strong (general)
WTS.

\section{Presetting}

For physical simplicity, we will use SI unit and Coulomb gauge $\nabla \cdot
\vec{A}=0$ in this paper. The scalar potential takes the form $\varphi
\left( \vec{x},t\right) =\int \frac{\rho _{e}\left( \vec{x}^{\prime
},t\right) }{4\pi \varepsilon _{0}\left\vert \vec{x}-\vec{x}^{\prime
}\right\vert }d^{3}\vec{x}^{\prime }$ under Coulomb gauge condition. Since
there is no net free charge $\rho _{e}$ in plasma, we have $\varphi \left(
\vec{x},t\right) \equiv 0$.

The vector potential $\vec{A}\left( \vec{x},t\right) $ related to a uniform
magnetic field $\vec{B}_{0}\left( t\right) $ is unbounded in space,
therefore we can not expand such $\vec{A}$ into Fourier series. Fortunately,
uniform magnetic field $\vec{B}_{0}$ (in full space, or in a sufficiently
large volume $V$) does not exist in real physical world. Moreover, we can
safely assume that $\vec{A}$ and $\vec{B}$ both vanish at the boundary $%
\partial V$ of $V$. In the following discussion, we assume that all the
fields $\vec{A}$, $\vec{B}$, $\vec{E}$, $\vec{J}$, $\vec{v}$\ are "good
enough" so that they can be expanded into Fourier series as

\begin{equation}
\left( \vec{A},\vec{B},\vec{E},\vec{J},\vec{v}\right) =\sum\limits_{\vec{k}%
\neq 0}\left( \vec{A}_{\vec{k}},\vec{B}_{\vec{k}},\vec{E}_{\vec{k}},\vec{J}_{%
\vec{k}},\vec{v}_{\vec{k}}\right) e^{i\vec{k}\cdot \vec{x}}.  \label{2}
\end{equation}

\bigskip For a finite volume $V$, the magnitude of nonzero $\vec{k}$ in Eq. (%
\ref{2}) has a universal lower limit $k_{0}>0$. $k_{0}$ depends on the size
and shape of $V$ and the boundary conditions only. For example, let $V$ be a
cubic with side length $L$ and choose the boundary conditions so that the
fields vanishes at the boundary, then $k_{0}=\frac{\pi }{L}$.

The inner product $\left\langle \vec{X},\vec{Y}\right\rangle $ of two real
fields $\vec{X}$, $\vec{Y}$ in $V$ is defined as $\left\langle \vec{X},\vec{Y%
}\right\rangle =\int_{V}\vec{X}\cdot \vec{Y}d^{3}\vec{x}$. For any real
field $\vec{X}$ we have $\left( \vec{X}_{\vec{k}}\right) ^{\ast }=\vec{X}_{-%
\vec{k}}$. The inner product becomes%
\begin{equation}
\begin{array}{c}
\left\langle \vec{X},\vec{Y}\right\rangle =\int_{V}\left( \sum\limits_{\vec{k%
}\neq 0}\vec{X}_{\vec{k}}e^{i\vec{k}\cdot \vec{x}}\right) \cdot \left(
\sum\limits_{\vec{l}\neq 0}\vec{Y}_{\vec{l}}e^{i\vec{l}\cdot \vec{x}}\right)
d^{3}\vec{x} \\
=V\sum\limits_{\vec{k}\neq 0}\vec{X}_{\vec{k}}\cdot \vec{Y}_{\vec{k}}^{\ast
}=V\sum\limits_{\vec{k}\neq 0}\vec{X}_{\vec{k}}^{\ast }\cdot \vec{Y}_{\vec{k}%
}.%
\end{array}
\label{3}
\end{equation}

By Cauchy-Schwartz inequality, $\Delta _{\vec{X},\vec{Y}}=\left\langle \vec{X%
},\vec{X}\right\rangle \left\langle \vec{Y},\vec{Y}\right\rangle
-\left\langle \vec{X},\vec{Y}\right\rangle ^{2}\geqslant 0$, $\Delta _{\vec{X%
},\vec{Y}}=0$ if and only if $\vec{X}=\alpha \vec{Y}$, where $\alpha $ is a
constant. Therefore $\Delta _{\vec{X},\vec{Y}}$ does measure the deviation
of the two fields to the equation $\vec{X}=\alpha \vec{Y}$.

\section{Evolution of the Magnetic Field}

We have the following equations in real space and in the Fourier component
space%
\begin{equation}
\nabla \cdot \vec{E}=0,\ \vec{k}\cdot \vec{E}_{\vec{k}}=0.  \label{4}
\end{equation}%
\begin{equation}
\nabla \cdot \vec{B}=0,\ \vec{k}\cdot \vec{B}_{\vec{k}}=0.  \label{5}
\end{equation}%
Taking time derivative on Eq. (\ref{5}), we have$\ $%
\begin{equation}
\vec{k}\cdot \vec{B}_{\vec{k}}^{\left( n\right) }\equiv \vec{k}\cdot \frac{%
d^{n}}{dt^{n}}\vec{B}_{\vec{k}}\equiv 0.  \label{6}
\end{equation}%
Equations $\vec{B}=\nabla \times \vec{A}$ and $\nabla \cdot \vec{A}=0$ give
\begin{equation}
\vec{A}_{\vec{k}}=\frac{i\vec{k}\times \vec{B}_{\vec{k}}}{k^{2}}.  \label{7}
\end{equation}%
Faraday's law reads%
\begin{equation}
\nabla \times \vec{E}=-\frac{\partial \vec{B}}{\partial t},i\vec{k}\times
\vec{E}_{\vec{k}}=-\frac{d}{dt}\vec{B}_{\vec{k}}.  \label{8}
\end{equation}

\bigskip By Eq. (\ref{4}), we can solve $\vec{E}_{\vec{k}}$ from Eq. (\ref{8}%
)
\begin{equation}
\vec{E}_{\vec{k}}=-\frac{i\vec{k}\times \frac{d}{dt}\vec{B}_{\vec{k}}}{k^{2}}%
.  \label{9}
\end{equation}%
We also have Ohm's law%
\begin{equation}
\begin{array}{c}
\vec{J}=\sigma \left( \vec{E}+\vec{v}\times \vec{B}\right) , \\
\vec{J}_{\vec{k}}=\sigma \left( \vec{E}_{\vec{k}}+\sum\limits_{\vec{l}\neq 0,%
\vec{l}\neq \vec{k}}\vec{v}_{\vec{l}}\times \vec{B}_{\vec{k}-\vec{l}}\right)
\\
=\sigma \left( -\frac{i\vec{k}\times \frac{d}{dt}\vec{B}_{\vec{k}}}{k^{2}}%
+\sum\limits_{\vec{l}\neq 0,\vec{l}\neq \vec{k}}\vec{v}_{\vec{l}}\times \vec{%
B}_{\vec{k}-\vec{l}}\right) ,%
\end{array}
\label{10}
\end{equation}%
where $\sigma $ is the conductivity of the plasma. For simplicity we assume
that $\sigma $ is a constant, as what Qin et al did $\textsuperscript{%
\cite{14}}$. Let $\lambda $ and $\tau $ be the typical scale of field
variation in space and time respectively, in the case that $v\ll \frac{%
\lambda }{\tau }$, we have%
\begin{equation}
\vec{J}_{\vec{k}}\approx -\frac{i\sigma \vec{k}\times \frac{d}{dt}\vec{B}_{%
\vec{k}}}{k^{2}}.  \label{11}
\end{equation}

Equation $\nabla \times \vec{B}=\mu _{0}\vec{J}+\varepsilon _{0}\mu _{0}%
\frac{\partial \vec{E}}{\partial t}$ gives

\begin{equation}
\vec{J}_{\vec{k}}=\frac{i\vec{k}\times \vec{B}_{\vec{k}}}{\mu _{0}}%
+\varepsilon _{0}\frac{i\vec{k}\times \frac{d^{2}}{dt^{2}}\vec{B}_{\vec{k}}}{%
k^{2}}.  \label{12}
\end{equation}%
When the fields are varying slowly, i.e., $\frac{\lambda }{\tau }\ll c$, we
have%
\begin{equation}
\vec{J}_{\vec{k}}\approx \frac{i\vec{k}\times \vec{B}_{\vec{k}}}{\mu _{0}}.
\label{13}
\end{equation}

\bigskip Combining Eq. (\ref{6}), (\ref{11}) and (\ref{13}) we have
\begin{equation}
\frac{d}{dt}\vec{B}_{\vec{k}}\approx -\frac{k^{2}}{\sigma \mu _{0}}\vec{B}_{%
\vec{k}}.  \label{14}
\end{equation}%
Therefore
\begin{equation}
\vec{B}_{\vec{k}}\left( t\right) =\vec{B}_{\vec{k}}\left( 0\right) \exp
\left( -\frac{k^{2}}{\sigma \mu _{0}}t\right) .  \label{15}
\end{equation}

We see that $\vec{B}_{\vec{k}}\left( t\right) $ goes to zero exponentially. $%
\vec{B}_{\vec{k}}\left( t\right) $ with larger $k^{2}$ varnishes faster.\
When $t\rightarrow +\infty $, $\vec{B}_{\vec{k}}$ with minimum $k^{2}$
dominates. The fine spatial wrinkles (short wavelength fluctuations) of $%
\vec{B}$ will be ironed out by Joule heat exponentially.

\section{Conditions for Approaching Strong WTS}

Now we are ready to calculate the evolution of $\theta _{\nabla \times \vec{B%
},\vec{B}}$. Firstly,

\begin{equation}
\begin{array}{c}
\left\langle \nabla \times \vec{B},\nabla \times \vec{B}\right\rangle
=V\sum\limits_{\vec{k}\neq 0}\left\vert i\vec{k}\times \vec{B}_{\vec{k}%
}\right\vert ^{2} \\
=V\sum\limits_{\vec{k}\neq 0}k^{2}\left\vert \vec{B}_{\vec{k}}\left(
0\right) \right\vert ^{2}\exp \left( -\frac{2k^{2}}{\sigma \mu _{0}}t\right)
,%
\end{array}
\label{16}
\end{equation}

\begin{equation}
\left\langle \vec{B},\vec{B}\right\rangle =V\sum\limits_{\vec{k}\neq
0}\left\vert \vec{B}_{\vec{k}}\left( 0\right) \right\vert ^{2}\exp \left( -%
\frac{2k^{2}}{\sigma \mu _{0}}t\right) ,  \label{17}
\end{equation}%
\begin{equation}
\begin{array}{c}
\left\langle \nabla \times \vec{B},\vec{B}\right\rangle =V\sum\limits_{\vec{k%
}\neq 0}\left[ i\vec{k}\times \vec{B}_{\vec{k}}\left( 0\right) \right] \cdot
\vec{B}_{\vec{k}}^{\ast }\left( 0\right) e^{-\frac{2k^{2}}{\sigma \mu _{0}}%
t}.%
\end{array}
\label{18}
\end{equation}

Let $k_{\min }\geqslant k_{0}>0$ be the minimum value of $\left\vert \vec{k}%
\right\vert $ with $\vec{B}_{\vec{k}}\left( 0\right) \neq 0$. When\bigskip\ $%
t\rightarrow +\infty $, terms with $\left\vert \vec{k}\right\vert =k_{\min }$
dominate in Eqs. (\ref{16}, \ref{17}, \ref{18}). We have
\begin{eqnarray}
&&\lim_{t\rightarrow +\infty }\cos \theta _{\nabla \times \vec{B},\vec{B}}
\label{19} \\
&=&\frac{\sum\limits_{\left\vert \vec{k}\right\vert =k_{\min }}\left[ i\vec{k%
}\times \vec{B}_{\vec{k}}\left( 0\right) \right] \cdot \vec{B}_{\vec{k}%
}^{\ast }\left( 0\right) }{k_{\min }\sum\limits_{\left\vert \vec{k}%
\right\vert =k_{\min }}\left\vert \vec{B}_{\vec{k}}\left( 0\right)
\right\vert ^{2}}  \nonumber \\
&\equiv &\cos \delta ,  \nonumber
\end{eqnarray}%
which is a constant depending on the initial spatial distribution of $\vec{B}
$ only. If $\sin \delta \neq 0$, plasma can never evolve to strong WTS. More
precisely, we have $\cos \theta _{\nabla \times \vec{B},\vec{B}}=\cos \delta
+O\left[ \exp \left( -\frac{2k_{next}^{2}-2k_{\min }^{2}}{\sigma \mu _{0}}%
t\right) \right] $, where $k_{next}$ is the minimum value of $\left\vert
\vec{k}\right\vert >k_{\min }$ with $\vec{B}_{\vec{k}}\left( 0\right) \neq 0$%
. We see that $\sin \theta _{\nabla \times \vec{B},\vec{B}}$ will go to $0$
at the rate $O\left[ \exp \left( -\frac{k_{next}^{2}-k_{\min }^{2}}{\sigma
\mu _{0}}t\right) \right] $ if $\sin \delta =0$. Recalling the fact that $%
\vec{B}_{\vec{k}}\left( t\right) =\vec{B}_{\vec{k}}\left( 0\right) \exp
\left( -\frac{k^{2}}{\sigma \mu _{0}}t\right) \leqslant \vec{B}_{\vec{k}%
}\left( 0\right) \exp \left( -\frac{k_{\min }^{2}}{\sigma \mu _{0}}t\right) $%
, we see that a reasonable requirement is $k_{next}^{2}>2k_{\min }^{2}$ so
that plasma can approach strong WTS \textquotedblleft
before\textquotedblright\ $\vec{B}$ practically goes to zero in the case $%
\sin \delta =0$.

As a by-product of Eq. (\ref{15}), we calculate%
\begin{equation}
\begin{array}{c}
\Delta =\left\langle \vec{A},\vec{A}\right\rangle \left\langle \vec{B},\vec{B%
}\right\rangle -\left\langle \vec{A},\vec{B}\right\rangle ^{2} \\
\leqslant \left\langle \vec{A},\vec{A}\right\rangle \left\langle \vec{B},%
\vec{B}\right\rangle \\
=V^{2}\left[ \sum\limits_{\vec{k}\neq 0}\frac{\left\vert \vec{B}_{\vec{k}%
}\left( 0\right) \right\vert ^{2}}{k^{2}}e^{-\frac{2k^{2}}{\sigma \mu _{0}}t}%
\right] \left[ \sum\limits_{\vec{k}\neq 0}\left\vert \vec{B}_{\vec{k}}\left(
0\right) \right\vert ^{2}e^{-\frac{2k^{2}}{\sigma \mu _{0}}t}\right] \\
\leqslant \frac{V^{2}}{k_{\min }^{2}}\left[ \sum\limits_{\left\vert \vec{k}%
\right\vert =k_{\min }}\left\vert \vec{B}_{\vec{k}}\left( 0\right)
\right\vert ^{2}\right] ^{2}e^{-\frac{4k_{\min }^{2}}{\sigma \mu _{0}}t},%
\end{array}
\label{20}
\end{equation}%
which does decreases exponentially with time at the rate $O\left[ \exp
\left( -\frac{4k_{\min }^{2}}{\sigma \mu _{0}}t\right) \right] $. But this
is merely a trivial inference of the fact that $\vec{B}$ vanishes at the
rate $O\left[ \exp \left( -\frac{k_{\min }^{2}}{\sigma \mu _{0}}t\right) %
\right] $.

If $\alpha $ in equation $\nabla \times \vec{B}=\alpha \vec{B}$ is not a
constant, as is widely discussed, the plasma is in general WTS. In this case
$\sin \theta _{\nabla \times \vec{B},\vec{B}}$ is generally nonzero,
therefore $\theta _{\nabla \times \vec{B},\vec{B}}$ can not measure the
deviation of plasma to the general WTS. A better quantity that measures the
deviation of plasma to the general WTS is the relative residual error $R$
defined by

\begin{equation}
R^{2}=\frac{\left\langle \left( \nabla \times \vec{B}\right) \times \vec{B}%
,\left( \nabla \times \vec{B}\right) \times \vec{B}\right\rangle }{k_{\min
}^{2}\left\langle \vec{B},\vec{B}\right\rangle ^{2}}.  \label{21}
\end{equation}

\bigskip It is obvious that $R=0$ if and only if $\left( \nabla \times \vec{B%
}\right) \times \vec{B}\equiv 0$ in $V$, which means $\nabla \times \vec{B}%
=\alpha \left( \vec{x}\right) \vec{B}$.

We have%
\begin{equation}
\left[ \left( \nabla \times \vec{B}\right) \times \vec{B}\right] _{\vec{k}%
}=\sum\limits_{\vec{l}\neq 0,\ \vec{l}\neq \vec{k}}\left( i\vec{l}\times
\vec{B}_{\vec{l}}\right) \times \vec{B}_{\vec{k}-\vec{l}},  \label{22}
\end{equation}

therefore%
\begin{equation}
\begin{array}{c}
R^{2}=\frac{\sum\limits_{\vec{k}\neq 0}\left\vert \sum\limits_{\vec{l}\neq
0,\ \vec{l}\neq \vec{k}}\left( i\vec{l}\times \vec{B}_{\vec{l}}\left(
0\right) \right) \times \vec{B}_{\vec{k}-\vec{l}}\left( 0\right) e^{-\frac{%
l^{2}+\left( \vec{k}-\vec{l}\right) ^{2}}{\sigma \mu _{0}}t}\right\vert ^{2}%
}{k_{\min }^{2}\left( \sum\limits_{\vec{k}\neq 0}\vec{B}_{\vec{k}}\left(
0\right) \cdot \vec{B}_{\vec{k}}^{\ast }\left( 0\right) e^{-\frac{2k^{2}}{%
\sigma \mu _{0}}t}\right) ^{2}} \\
\rightarrow \frac{\sum\limits_{\vec{k}\neq 0}\left\vert
\sum\limits_{\left\vert \vec{l}\right\vert =\left\vert \vec{k}-\vec{l}%
\right\vert =k_{\min }}\left( \vec{l}\times \vec{B}_{\vec{l}}\left( 0\right)
\right) \times \vec{B}_{\vec{k}-\vec{l}}\left( 0\right) \right\vert ^{2}}{%
k_{\min }^{2}\left( \sum\limits_{\left\vert \vec{k}\right\vert =k_{\min }}%
\vec{B}_{\vec{k}}\left( 0\right) \cdot \vec{B}_{\vec{k}}^{\ast }\left(
0\right) \right) ^{2}}%
\end{array}
\label{23}
\end{equation}%
as $t\rightarrow \infty $, which is also a constant and nonzero in general
cases.

\section{Further Discussions}

In the most general cases, $W$ will decay at the rate%
\begin{equation}
\begin{array}{c}
\frac{dW}{dt}=2\left\langle \frac{\partial \vec{B}}{\partial t},\vec{B}%
\right\rangle =-2\left\langle \nabla \times \vec{E},\vec{B}\right\rangle  \\
=2\int_{V}\nabla \cdot \left( \vec{B}\times \vec{E}\right) d^{3}\vec{x}%
-2\left\langle \vec{E},\nabla \times \vec{B}\right\rangle  \\
=2\int_{\partial V}\left( \vec{B}\times \vec{E}\right) \cdot d\vec{S}%
-2\left\langle \vec{E},\mu _{0}\vec{J}+\varepsilon _{0}\mu _{0}\frac{%
\partial \vec{E}}{\partial t}\right\rangle  \\
=-2\varepsilon _{0}\mu _{0}\left\langle \vec{E},\frac{\partial \vec{E}}{%
\partial t}\right\rangle -2\mu _{0}\left\langle \vec{E},\vec{J}\right\rangle
\\
\approx 2\mu _{0}\left\langle \vec{v}\times \vec{B},\vec{J}\right\rangle
-\left\langle \frac{2\mu _{0}}{\sigma }\vec{J},\vec{J}\right\rangle  \\
\approx -\left\langle \frac{2\mu _{0}}{\sigma }\vec{J},\vec{J}\right\rangle
\leqslant -\frac{2\mu _{0}}{\sigma _{\max }}\left\langle \vec{J},\vec{J}%
\right\rangle  \\
\leqslant -\frac{2\mu _{0}k_{\min }^{2}}{\sigma _{\max }}W.%
\end{array}
\label{24}
\end{equation}%
where $\sigma _{\max }=\sup\limits_{\vec{x}\in V}\left[ \sigma \left( \vec{x}%
\right) \right] <\infty $. We see that $\vec{B}$ will varnish due to Joule
heat within time scale $\frac{\sigma _{\max }}{\mu _{0}k_{\min }^{2}}$ for
any real plasma with finite conductivity $\sigma \left( \vec{x}\right) $.
The adjective "strong" is necessary when talking about WTS in physics, and
the corresponding time scale is indispensable when talking about the
evolution towards WTS, since $\lim\limits_{t\rightarrow +\infty }\vec{B}=0$.

In the above discussions we see that under the assumptions we have made, the
WTS is reached only if the initial distribution of $\vec{B}$ satisfies some
constrains. But astrophysical and laboratory observations indicate that WTS
is the general status for plasma. The only significant factor that we have
not taken into account above is the non-homogeneousity of $\sigma $.

For real plasma with inhomogeneous $\sigma =\sigma _{0}+\sum\limits_{\vec{k}%
\neq 0}\sigma _{\vec{k}}e^{i\vec{k}\cdot \vec{x}}$, Eq. (\ref{11}) becomes $%
\vec{J}_{\vec{k}}\approx -\sigma _{0}\frac{i\vec{k}\times \frac{d}{dt}\vec{B}%
_{\vec{k}}}{k^{2}}-\sum\limits_{\vec{l}\neq 0}\sigma _{\vec{l}}\frac{i\left(
\vec{k}-\vec{l}\right) \times \frac{d}{dt}\vec{B}_{\vec{k}-\vec{l}}}{\left(
\vec{k}-\vec{l}\right) ^{2}}$. Therefore Eq. (\ref{14}) becomes%
\begin{equation}
\begin{array}{c}
\frac{d}{dt}\vec{B}_{\vec{k}}\approx -\frac{k^{2}\vec{B}_{\vec{k}}}{\sigma
_{0}\mu _{0}} \\
+\sum\limits_{\vec{l}\neq 0,\vec{l}\neq \vec{k}}\frac{\sigma _{\vec{l}}}{%
\sigma _{0}}\frac{\vec{k}\times \left[ \left( \vec{k}-\vec{l}\right) \times
\frac{d}{dt}\vec{B}_{\vec{k}-\vec{l}}\right] }{\left( \vec{k}-\vec{l}\right)
^{2}}.%
\end{array}
\label{25}
\end{equation}%
Plug the first order approximation $\vec{B}_{\vec{k}}\left( t\right) \approx
\vec{B}_{\vec{k}}\left( 0\right) e^{-\frac{k^{2}}{\sigma _{0}\mu _{0}}t}$ in
the last term in Eq. (\ref{25}), we have%
\begin{equation}
\begin{array}{c}
\frac{d}{dt}\vec{B}_{\vec{k}}\approx -\frac{k^{2}\vec{B}_{\vec{k}}}{\sigma
_{0}\mu _{0}} \\
-\sum\limits_{\vec{l}\neq 0,\vec{l}\neq \vec{k}}\frac{\sigma _{\vec{l}}}{%
\sigma _{0}}\frac{\vec{k}\times \left[ \left( \vec{k}-\vec{l}\right) \times
\vec{B}_{\vec{k}-\vec{l}}\left( 0\right) \right] }{\sigma _{0}\mu _{0}}e^{-%
\frac{\left( \vec{k}-\vec{l}\right) ^{2}}{\sigma _{0}\mu _{0}}t}.%
\end{array}
\label{26}
\end{equation}
We see that even for $k>k_{\min }$, there could still exist some $\vec{l}%
\neq 0$ so that $\left\vert \vec{k}-\vec{l}\right\vert =k_{\min }<k$, which
will generate one term proportional to $e^{-\frac{k_{\min }^{2}}{\sigma
_{0}\mu _{0}}t}$ that dominates in $\vec{B}_{\vec{k}}$ when $t\rightarrow
+\infty $ and change the values of $\lim\limits_{t\rightarrow +\infty
}\theta _{\nabla \times \vec{B},\vec{B}}$ and $\lim\limits_{t\rightarrow
+\infty }R$. Thus the non-homogeneousity of $\sigma $ is one very possible
candidate for the real power that pushes plasma towards (general) WTS before
$\vec{B}$ practically goes to zero.

\bigskip

\end{document}